\PassOptionsToPackage{numbers,sort&compress}{natbib} 
\documentclass[%
preprint,
titlepage,
superscriptaddress,
nofootinbib,
 amsmath,amssymb,
 aps,
]{revtex4-2}

\usepackage{graphicx}
\usepackage{dcolumn}
\usepackage{bm}
\usepackage{threeparttable}
\usepackage{caption}
\usepackage{xcolor}
\usepackage[colorlinks=true, citecolor=blue, linkcolor=blue, urlcolor=blue]{hyperref}
\usepackage[section]{placeins}
\begin{document}

\preprint{}

\title{
    Higher-Order Analytical Expansion of Thawing Dark Energy with an Exponential Potential 
}
\author{Naoto Maki}
\affiliation{Department of Astronomical Science, The Graduate University for Advanced Studies (SOKENDAI), 2-21-1 Osawa, Mitaka, Tokyo 181-8588, Japan}
\affiliation{Division of Science, National Astronomical Observatory of Japan, 2-21-1 Osawa, Mitaka, Tokyo 181-8588, Japan}

\author{Kazunori Kohri}
\affiliation{Division of Science, National Astronomical Observatory of Japan, 2-21-1 Osawa, Mitaka, Tokyo 181-8588, Japan}
\affiliation{Department of Astronomy, The University of Tokyo, Bunkyo-ku, Hongo, Tokyo 113-0033, Japan}
\affiliation{Department of Astronomical Science, The Graduate University for Advanced Studies (SOKENDAI), 2-21-1 Osawa, Mitaka, Tokyo 181-8588, Japan}
\affiliation{Theory Center, IPNS, KEK,
1-1 Oho, Tsukuba, Ibaraki 305-0801, Japan}
\affiliation{Kavli IPMU (WPI), UTIAS, The University of Tokyo, Kashiwa, Chiba 277-8583, Japan}

\date{\today}

\begin{abstract}
    Motivated by recent DESI results suggesting dynamical dark energy, we investigate the thawing scenario in quintessence with an exponential potential, $V=V_0e^{-\lambda\phi/m_{\mathrm{pl}}}$, by analytically expanding the deviation of the equation of state parameter $w_{\phi}$ from $-1$ in powers of $\lambda$. In addition to the previously known leading-order result at $\mathcal{O}(\lambda^2)$, we derive the $\mathcal{O}(\lambda^4)$ correction as a function of the density parameter $\Omega_\phi$. We show that a consistent determination of the redshift dependence of $w_\phi$ through $\mathcal{O}(\lambda^4)$ requires corrections to the background expansion. We obtain the required correction by expanding $\Omega_\phi$ in powers of $\lambda$ around its $\Lambda\mathrm{CDM}$ value. Comparison with numerical solutions demonstrates that the $\mathcal{O}(\lambda^4)$ expansion provides a more accurate approximation than the leading-order result. Our analytical approximation, which consistently incorporates the $\mathcal{O}(\lambda^4)$ correction, will provide a potentially useful tool for distinguishing the exponential quintessence model from other dark energy models in future observations.
\end{abstract}

\maketitle

\section{Introduction}
Dark energy is an unknown energy component required to explain the accelerating expansion of the universe. In the standard cosmological model known as the $\Lambda\mathrm{CDM}$ model, this accelerated expansion is realized by a positive cosmological constant characterized by a constant energy density, yielding an equation of state parameter $w_{\Lambda}=-1$.

Recently, however, hints of time variation in the equation of state parameter, which cannot be explained by such a $\Lambda\mathrm{CDM}$ model, have been indicated by DESI \cite{DESI:2024mwx,DESI:2025zgx,DESI:2025fii}. For various interpretations and discussions regarding DESI, see Refs.~\cite{Tada:2024znt,Yang:2024kdo,Gialamas:2024lyw,RoyChoudhury:2025iis,Maqsood:2026krg,Woo:2026ice,Colgain:2025nzf,Li:2026asg}. One of the most representative theoretical models describing such time-varying dark energy is quintessence, which drives the accelerated expansion via a canonical scalar field $\phi$ with a potential $V(\phi)$. Although quintessence with a non-negative potential satisfies $-1\leq w_\phi\leq1$ and cannot explain phantom behavior, it serves as the most fundamental benchmark among dynamical dark energy models. Therefore, refining its theoretical foundations and improving the precision of its predictions is important.

Among quintessence models, the class that behaves like a cosmological constant in the past and deviates from this behavior near the present is called thawing quintessence \cite{Caldwell:2005tm,Linder:2006sv}. Among the potentials that yield solutions exhibiting thawing behavior, models with an exponential potential $V(\phi)=V_0e^{-\lambda\phi/m_{\mathrm{pl}}}$ (where $m_{\mathrm{pl}}$ is the reduced Planck mass, and $V_0$ and $\lambda$ are constant parameters) are motivated by string theory \cite{Obied:2018sgi,Cicoli:2023opf,Nitta:2025mpi}, and have been widely studied \cite{Halliwell:1986ja,Burd:1988ss,Ferreira:1997hj,Copeland:1997et,vandenHoogen:1999qq,Kolda:2001ex,Copeland:2006wr,Boehmer:2010jqg,Urena-Lopez:2011gxx,Tamanini:2014mpa,Gosenca:2015qha,Bahamonde:2017ize,SavasArapoglu:2017pyh,Andriot:2024jsh,Andriot:2026lac,Andriot:2024sif,Maki:2026zrn,Tsimpis:2026nyn}. They have also been discussed in the context of modified gravity \cite{Sadatian:2024oab,Hong:2025tyi,Devi:2026dwg,Pookkillath:2026wyg}, quantum cosmology \cite{Ratra:1989uv,Rebesh:2019pbw,Maki:2026fgm}, multi-field extensions \cite{Liddle:1998jc,Malik:1998gy,Chiba:2014sda,Alestas:2025syk}, and extensions to interacting or non-minimally coupled models \cite{Nozari:2016ilx,Pourtsidou:2025sdd,BeltranJimenez:2026ymd,Wang:2026gvg,Tzanni:2014eja,Wang:2026wrk}. Furthermore, a relaxation mechanism for the fine-tuning problem using a kination phase in this model has also been discussed recently \cite{Maki:2026qpf}. For observational constraints, see Refs.~\cite{Bhattacharya:2024hep,Ramadan:2024kmn,Akrami:2025zlb,Bayat:2025xfr,Pourtsidou:2025sdd,BeltranJimenez:2026ymd,Sultana:2026ych,SanchezLopez:2025uzw}.

A recent theoretical development in this exponential quintessence model is the finding that the present equation of state parameter $w_{\phi0}$ and the potential parameter $\lambda$ are restricted to a very narrow region in the parameter space in order to realize sufficiently long matter and radiation dominated eras. This behavior was initially discovered numerically \cite{Andriot:2024jsh} (see also \cite{Andriot:2026lac,Andriot:2024sif}), and later explained analytically by Ref.~\cite{Maki:2026zrn} using a method that expands the deviation of the equation of state parameter $w_\phi$ from $-1$ in terms of the density parameter $\Omega_\phi$. 
In this expansion, the solution is chosen such that $w_{\phi}\to -1$ as $\Omega_\phi\to 0$.
Namely, the solution obtained from this expansion corresponds to the thawing branch, where a scalar field that is frozen in the early universe starts to roll once its energy density becomes cosmologically relevant. As another analytical description of such thawing solutions, Ref.~\cite{Scherrer:2007pu} approximated the equations of $1+w_{\phi}$ for models where the parameter $\lambda=-m_{\mathrm{pl}}\frac{V_{,\phi}}{V}$ (where $V_{,\phi}=\frac{dV}{d\phi}$) is nearly constant. This framework was extended to cases with non-minimal coupling and phantom fields \cite{Scherrer:2008be,Gupta:2009kk}, spatial curvature \cite{delCampo:2010fg}, and non-canonical kinetic terms \cite{Li:2016grl}. More recently, it was revisited in the context of cosmography \cite{Chakraborty:2026ooz}. 
The connection with the approach of Ref.~\cite{Maki:2026zrn} becomes clear when the solution in Ref.~\cite{Scherrer:2007pu} is expressed as an expansion in $\lambda$. In fact, it corresponds precisely to the leading $\mathcal{O}(\lambda^2)$ term in the expansion of $1+w_\phi$. Therefore, Ref.~\cite{Maki:2026zrn} and Ref.~\cite{Scherrer:2007pu} represent the same thawing branch through different expansions; the former expands the deviation of $w_\phi$ from $-1$ in terms of $\Omega_\phi$, while the latter investigates the leading-order term of the $\lambda$ expansion. Recently, a complementary analytical approach for general thawing quintessence based on a Pad\'e approximation has also been developed in Ref.~\cite{Alho:2024qds}.

In this study, we limit our focus to the exponential potential, where $\lambda=-m_{\mathrm{pl}}\frac{V_{,\phi}}{V}=\mathrm{const}$ holds exactly, and extend the framework of Ref.~\cite{Scherrer:2007pu} to the next higher order, $\mathcal{O}(\lambda^4)$. We also demonstrate that the obtained results reproduce the general expression for the coefficients in Ref.~\cite{Maki:2026zrn}. In Ref.~\cite{Scherrer:2007pu}, the redshift dependence of $w_\phi$ is evaluated on the $\Lambda\mathrm{CDM}$ background (i.e., $\Omega_\phi$ is approximated by that of the $\Lambda\mathrm{CDM}$ model). Although this approximation is consistent up to $\mathcal{O}(\lambda^2)$, a calculation up to $\mathcal{O}(\lambda^4)$ must also include the correction to the background expansion induced by the scalar-field dynamics. Taking this correction into account, we derive $w_\phi(z)$ consistently up to $\mathcal{O}(\lambda^4)$. Finally, we eliminate $\lambda$ from the expansion of $w_\phi(z)$ and obtain an expression in terms of the present equation of state parameter $w_{\phi0}$ and the present density parameter $\Omega_{\phi0}$.

This paper is organized as follows. In Sec.~\ref{sec:formalism}, we formulate the basic cosmological equations. In Sec.~\ref{sec:analytical_expansion}, we introduce two different expansions of $w_\phi$ and derive the previously unknown higher-order terms for the $\lambda$ expansion. In Sec.~\ref{sec:redshift_dependence}, we derive the redshift dependence of $w_\phi$ up to $\mathcal{O}(\lambda^4)$. Finally, Sec.~\ref{sec:conclusion} is devoted to our conclusions.

\section{Formalism}\label{sec:formalism}
We consider a spatially flat, homogeneous, and isotropic universe containing pressureless matter and a canonical scalar field. The energy density and pressure of the scalar field are given by 
\begin{align}
    \rho_\phi&=\frac{1}{2}\dot{\phi}^2+V(\phi),\\
    p_\phi&=\frac{1}{2}\dot{\phi}^2-V(\phi),
\end{align}
respectively. The equation of state parameter of the scalar field is defined as
\begin{align}
    w_\phi\equiv\frac{p_\phi}{\rho_\phi}=\frac{\frac{1}{2}\dot{\phi}^2-V(\phi)}{\frac{1}{2}\dot{\phi}^2+V(\phi)}.
\end{align}
The equation of motion for the scalar field is 
\begin{align}
    \ddot{\phi}+3H\dot{\phi}+V_{,\phi}=0,
\end{align}
where $H\equiv\frac{\dot{a}}{a}$ is the Hubble parameter, an overdot denotes differentiation with respect to cosmic time, and $V_{,\phi}\equiv \frac{dV}{d\phi}$. The Friedmann equations are given by
\begin{align}
  &3m_{\mathrm{pl}}^2 H^2 = \frac{1}{2}\dot{\phi}^2+V(\phi)+\rho_m,\\
  &2m_{\mathrm{pl}}^2 \dot{H}=-\dot{\phi}^2-\rho_m\label{acceleration_equation},
\end{align}
where $\rho_m$ is the energy density of pressureless matter and $m_{\mathrm{pl}}$ is the reduced Planck mass. The density parameters are defined as 
\begin{align}
    \Omega_\phi\equiv\frac{\rho_\phi}{3m_{\mathrm{pl}}^2H^2}, \qquad \Omega_m\equiv\frac{\rho_m}{3m_{\mathrm{pl}}^2H^2}.
\end{align}
Therefore, we have $\Omega_m+\Omega_\phi=1$. Furthermore, assuming $\dot{\phi}>0$, we obtain
\begin{align}
    \frac{dw_\phi}{d\ln a}=-(1-w_\phi)\left\{ 3(1+w_\phi)-\lambda\sqrt{3\Omega_\phi(1+w_\phi)} \right\},
\end{align}
where
\begin{align}
    \lambda=-m_{\mathrm{pl}}\frac{V_{,\phi}}{V}.
\end{align}
The parameter $\lambda$ is related to the potential slow-roll parameter commonly used in the context of inflation, $\epsilon_V\equiv\frac{m_{\mathrm{pl}}^2}{2}\left(V_{,\phi}/V\right)^2$, through the relation $\epsilon_V=\lambda^2/2$. For a universe containing pressureless matter and the scalar field, we also obtain
\begin{align}
    \frac{d\Omega_\phi}{d\ln a}&=-3w_\phi(1-\Omega_\phi)\Omega_\phi.
    \label{eq:domega_dlna}
\end{align}
Combining these two equations, we obtain 
\begin{align}
    \frac{dw_\phi}{d\Omega_\phi}=\frac{(1-w_\phi)\left\{ 3(1+w_\phi)-\lambda\sqrt{3\Omega_\phi(1+w_\phi)} \right\}}{3w_\phi\Omega_\phi(1-\Omega_\phi)}.
    \label{eq:dwdomega}
\end{align}
When $\lambda$ is constant, the potential takes the 
exponential form
\begin{align}
    V=V_0e^{-\lambda\phi/m_{\mathrm{pl}}},
\end{align}
where $V_0$ is a constant. Throughout this work, we consider an exponential potential, corresponding to $\lambda=\mathrm{const}$, and assume $\lambda>0$ and $V_0>0$.

\section{Analytical Expansion}
\label{sec:analytical_expansion}
In this section, we consider a scenario in which $w_\phi$ begins
to deviate from $-1$ at late times as the scalar field starts to roll. The class of models exhibiting this behavior is referred to as thawing quintessence. For the purposes of this study, we consider two different approaches for expanding this deviation. The first approach is to expand $w_\phi$ in terms of $\Omega_\phi$ as $w_\phi=-1+c_1(\lambda)\Omega_\phi+c_2(\lambda)\Omega_\phi^2+\cdots$. This method was analyzed in Ref.~\cite{Maki:2026zrn}. The second is to expand $w_\phi$ in terms of $\lambda$ as $w_\phi=-1+f_2(\Omega_\phi)\lambda^2+f_4(\Omega_\phi)\lambda^4+\cdots$.
The leading-order coefficient $f_2(\Omega_\phi)$ was analytically derived in Ref.~\cite{Scherrer:2007pu}. In this section, we mainly work on the second approach. We extend the result of Ref.~\cite{Scherrer:2007pu} up to the fourth order in $\lambda$ and analytically determine the coefficient $f_4(\Omega_\phi)$. Furthermore, we demonstrate that this result exactly reproduces the coefficients derived from the first method (i.e., expansion in $\Omega_\phi$).

First, let us expand $w_\phi$ around $-1$ in powers of $\Omega_\phi$ as 
\begin{align}
    1+w_\phi=\sum_{n=1}^{\infty}c_n \Omega_\phi^n .
\end{align}
Substituting this into Eq.~\eqref{eq:dwdomega} and comparing the coefficients of each power of $\Omega_\phi$, we obtain 
\begin{align}
    w_\phi=&-1+\frac{4}{27}\lambda^2\Omega_\phi+\left(\frac{16}{3645}\lambda^4+ \frac{8}{135}\lambda^2 \right)\Omega_\phi^2\notag\\
    &+\left( \frac{1712}{3444525}\lambda^6+\frac{352}{127575}\lambda^4+\frac{148}{4725}\lambda^2 \right) \Omega_\phi^3+\mathcal{O}(\Omega_\phi^4).
    \label{eq:expansion_in_omega}
\end{align}
This result was derived in Ref.~\cite{Maki:2026zrn}.
\footnote{
    Note that the first term, $\frac{4}{27}\lambda^2\Omega_\phi$, was obtained in Ref.~\cite{Cahn:2008gk} as the leading-order coefficient for general thawing quintessence.
} In Ref.~\cite{Maki:2026zrn}, it was confirmed that imposing the present boundary conditions such that this relation is satisfied guarantees the thawing behavior when solving the equations backward in time. The advantage of this expansion is that calculation of higher-order terms can be performed relatively easily through a mechanical procedure.
Next, we consider the expansion of $w_\phi$ around $-1$ in a power series of $\lambda$. In Eq.~\eqref{eq:expansion_in_omega}, odd powers of $\lambda$ do not appear. Furthermore, the recurrence relations do not yield odd powers of $\lambda$ at any order. Therefore, $1+w_\phi$ can be expanded using only even powers of $\lambda$ as follows:
\begin{align}
    1+w_\phi=\sum_{n=1}^{\infty}f_{2n}(\Omega_\phi)\lambda^{2n}.
\end{align}
Substituting this into Eq.~\eqref{eq:dwdomega} and equating the coefficient of $\lambda^2$, we obtain the following differential equation for $f_2$:
\begin{align}
    \frac{df_2}{d\Omega_\phi}=-\frac{2f_2}{\Omega_\phi(1-\Omega_\phi)}+\frac{2\sqrt{3\Omega_\phi f_2}}{3\Omega_\phi(1-\Omega_\phi)} .
\end{align}
By defining $f_2=\Theta^2$, this equation reduces to a linear differential equation:
\begin{align}
    \frac{d\Theta}{d\Omega_\phi}=-\frac{\Theta}{\Omega_\phi(1-\Omega_\phi)}+\frac{1}{\sqrt{3\Omega_\phi}(1-\Omega_\phi)} .
    \label{eq:f_2_equation}
\end{align}
Solving this equation, we obtain the general solution:
\begin{align}
    \Theta=\frac{1}{\sqrt{3}}\left\{ \frac{1}{\sqrt{\Omega_\phi}}-\frac{1}{2}\left( \frac{1}{\Omega_\phi}-1 \right)\ln\left( \frac{1+\sqrt{\Omega_\phi}}{1-\sqrt{\Omega_\phi}} \right) \right\}+C\frac{1-\Omega_\phi}{\Omega_\phi},
\end{align}
where $C$ is an integration constant.
The second term diverges as $\Omega_\phi\to 0$ for $C\neq0$. Imposing the thawing boundary condition $1+w_\phi\to 0$ as $\Omega_\phi\to 0$, we thus set $C=0$. Therefore, we obtain the solution satisfying the thawing boundary condition,
\begin{align}
    f_2(\Omega_\phi)=\Theta^2=\frac{1}{3}\left\{ \frac{1}{\sqrt{\Omega_\phi}}-\frac{1}{2}\left( \frac{1}{\Omega_\phi}-1 \right)\ln\left( \frac{1+\sqrt{\Omega_\phi}}{1-\sqrt{\Omega_\phi}} \right)\right\}^2,
\end{align}
which was previously obtained in Ref.~\cite{Scherrer:2007pu}. This can be expressed as a series expansion:
\begin{align}
    f_2(\Omega_\phi)&=\frac{4}{3}\sum_{k=1}^{\infty} \sum_{n=1}^{k}\frac{1}{4n^2-1}\frac{1}{4(k-n+1)^2-1} \Omega_\phi^k\notag\\
    &=\frac{4}{27}\Omega_\phi+\frac{8}{135}\Omega_\phi^2+\frac{148}{4725}\Omega_\phi^3+\cdots.
\end{align}
This is consistent with the coefficient of $\lambda^2$ in Eq.~\eqref{eq:expansion_in_omega}. Furthermore, the coefficient of $\lambda^4$ leads to the following equation.
\begin{align}
    \frac{df_4}{d\Omega_\phi}&=\frac{1}{3\Omega_\phi(1-\Omega_\phi)}\left( \sqrt{\frac{3\Omega_\phi}{f_2}}-6 \right)f_4+\frac{f_2^2}{\Omega_\phi(1-\Omega_\phi)}-\frac{f_2\sqrt{3\Omega_\phi f_2}}{3\Omega_\phi(1-\Omega_\phi)}+f_2\frac{df_2}{d\Omega_\phi}.
    \label{eq:f_4_equation}
\end{align}
This equation can be simplified to 
\begin{align}
    \frac{df_4}{d\Omega_\phi}&=\left\{ \frac{\Theta'}{\Theta}-\frac{1}{\Omega_\phi(1-\Omega_\phi)} \right\}f_4+\Theta^3\Theta',
\end{align}
where $\Theta=\sqrt{f_2}$ and $\Theta'=\frac{d\Theta}{d\Omega_\phi}$. This is also a linear differential equation and can be solved analytically (see Sec.~\ref{sec:f_4_derivation} for a detailed derivation):
\begin{align}
    f_4(\Omega_\phi)&=\frac{1-\Omega_\phi}{9\Omega_\phi}\left\{ \frac{1}{\sqrt{\Omega_\phi}}-\frac{1}{2}\left( \frac{1-\Omega_\phi}{\Omega_\phi} \right)\ln\left( \frac{1+\sqrt{\Omega_\phi}}{1-\sqrt{\Omega_\phi}} \right) \right\}\notag\\
    &\qquad\times\left\{ \frac{1}{2\sqrt{\Omega_\phi}}-\frac{3+\Omega_\phi}{4\Omega_\phi}\ln\left( \frac{1+\sqrt{\Omega_\phi}}{1-\sqrt{\Omega_\phi}} \right)\right.\notag\\
    &\qquad\qquad\left.+\frac{3-\Omega_\phi}{8\Omega_\phi^{3/2}}\ln^2\left( \frac{1+\sqrt{\Omega_\phi}}{1-\sqrt{\Omega_\phi}} \right)-\frac{(1-\Omega_\phi)^2}{16\Omega_\phi^2}\ln^3\left( \frac{1+\sqrt{\Omega_\phi}}{1-\sqrt{\Omega_\phi}} \right)\right\} .
    \label{eq:f_4}
\end{align}
The general solution contains an additional term $C(1-\Omega_\phi)\Theta/\Omega_\phi$, where $C$ is an integration constant. This term diverges for $\Omega_\phi\to0$ and is therefore excluded by the thawing boundary
condition. Eq.~\eqref{eq:f_4} is the solution obtained after setting $C=0$. Eq.~\eqref{eq:f_4} can be expressed as the following series expansion:
\begin{align}
    f_4(\Omega_\phi)&=\frac{16}{9}\left\{ C_1\Omega_\phi^2+\sum_{q=1}^{\infty}(C_{q+1}-C_q)\Omega_\phi^{q+2} \right\}\notag\\
    &=\frac{16}{3645}\Omega_\phi^2+\frac{352}{127575}\Omega_\phi^3+\mathcal{O}(\Omega_\phi^4)
    \label{eq:f_4_series_expansion}
\end{align}
where we define 
\begin{align}
    C_q\equiv\sum_{p=1}^{q}\sum_{l=1}^{p}\sum_{k=1}^{l}\sum_{m=1}^{k}\frac{1}{4(q-p+1)^2-1}\frac{1}{2p+3}\frac{1}{2(l-k)+3}\frac{1}{4m^2-1}\frac{1}{4(k-m+1)^2-1}.
\end{align}
It is straightforward to check that this agrees with the first few terms of the $\lambda^4$ coefficient in Eq.~\eqref{eq:expansion_in_omega}.

\section{Redshift Dependence of the Equation of State Parameter}
\label{sec:redshift_dependence}
\begin{figure}[p]
  \centering
  \includegraphics[scale = 0.6]{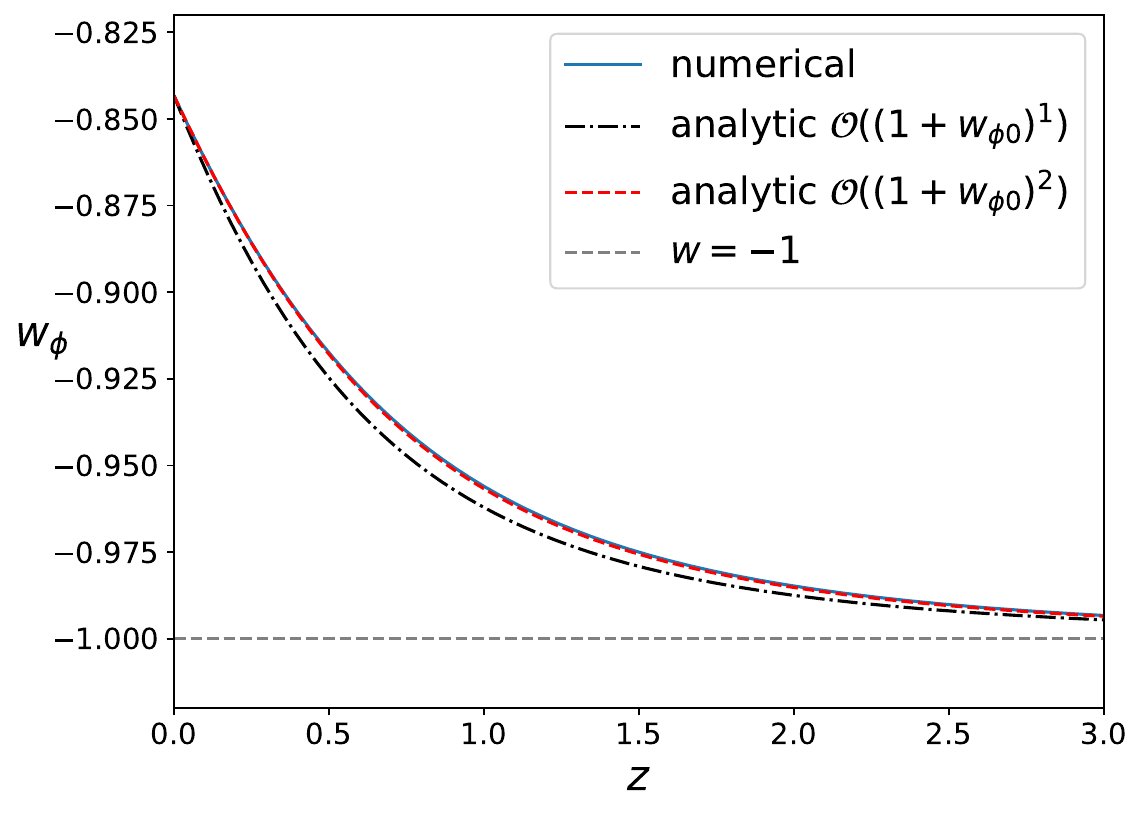}
  \caption{Comparison of the numerical evolution of $w_{\phi}(z)$ (solid blue) with the analytical expression in Eq.~\eqref{eq:w_z_w_phi0} truncated at first order (black dash-dotted) and second order (red dashed) in $(1+w_{\phi0})$. We set $\lambda=1$, $\Omega_{\phi0}=0.70$, and $w_{\phi0}=-0.843322$. The value of $w_{\phi0}$ is obtained by extending the expansion in Eq.~\eqref{eq:expansion_in_omega} to higher orders in $\Omega_\phi$. The horizontal grey dashed line denotes $w=-1$. }
  \label{fig:w_comparison}
\end{figure}

\begin{figure}[htbp]
  \centering
  \includegraphics[scale = 0.5]{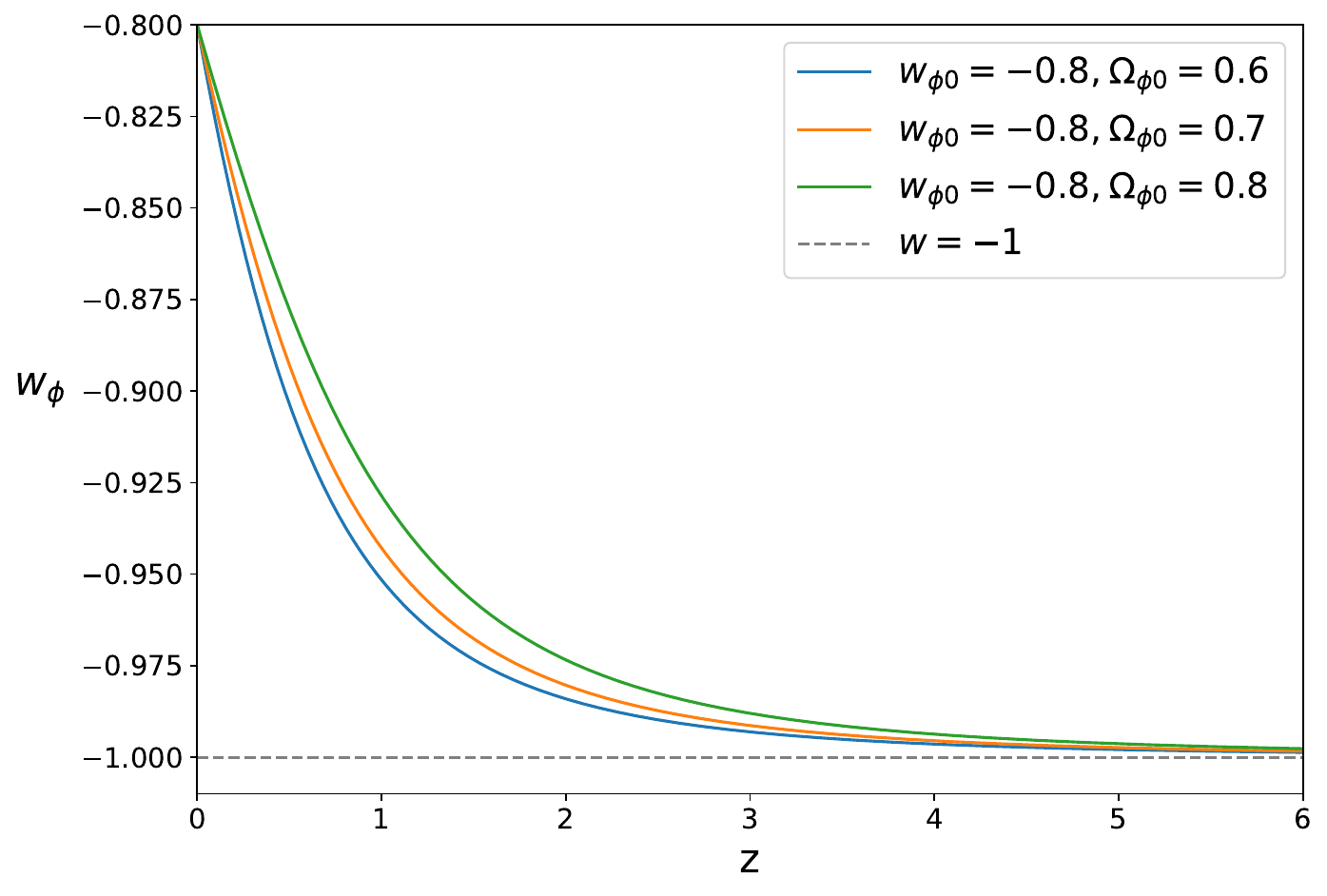}
  \caption{Redshift evolution of $w_\phi$ obtained from Eq.~\eqref{eq:w_z_w_phi0}, including terms up to second order in $(1+w_{\phi0})$, for $\Omega_{\phi0}=0.6$, $0.7$ and $0.8$. The present equation of state parameter is fixed at $w_{\phi0}=-0.8$. The horizontal grey dashed line denotes $w=-1$.}
  \label{fig:w_z_omega}
\end{figure}

\begin{figure}[htbp]
  \centering
  \includegraphics[scale = 0.5]{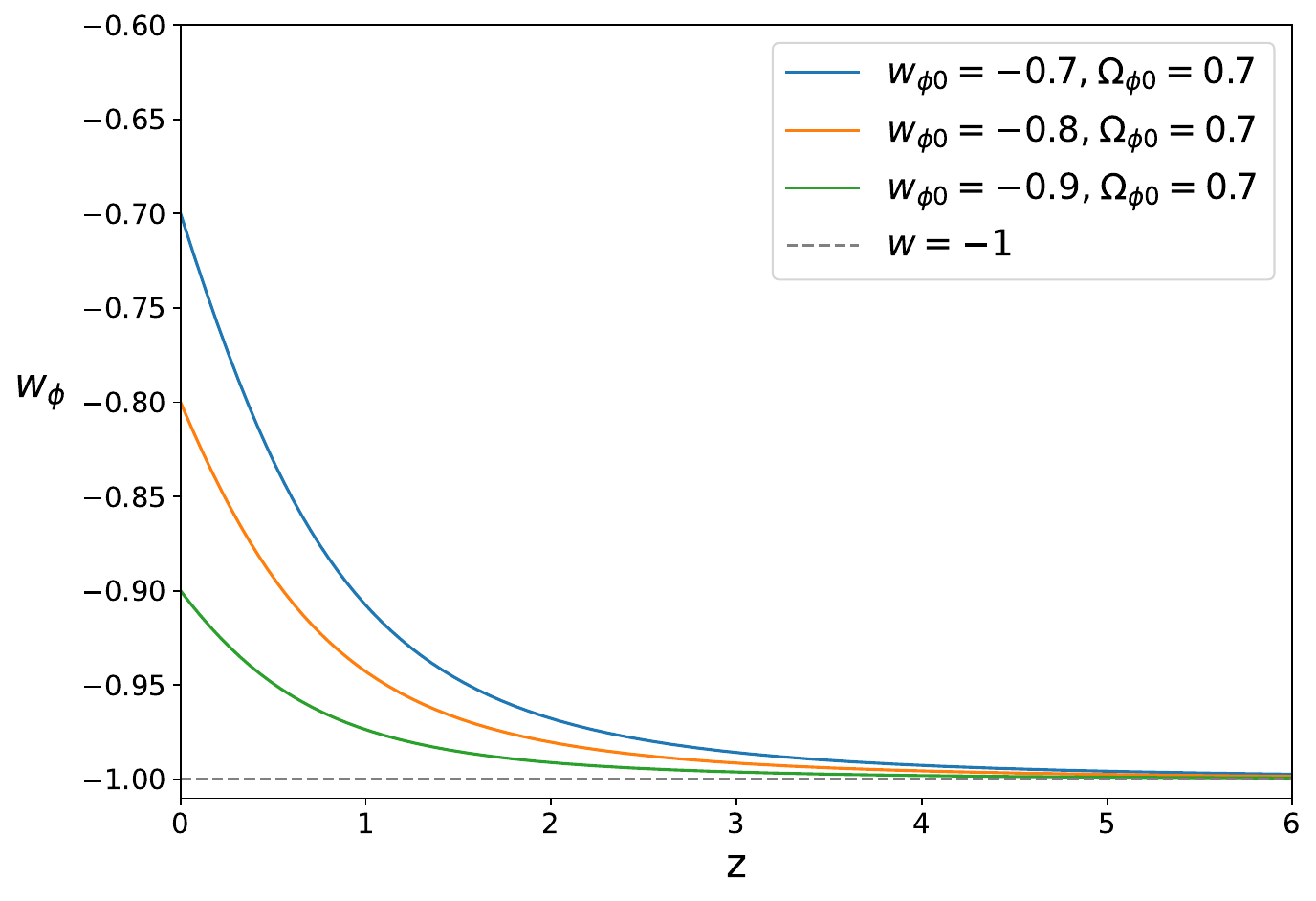}
  \caption{Redshift evolution of $w_\phi$ obtained from Eq.~\eqref{eq:w_z_w_phi0}, including terms up to second order in $(1+w_{\phi0})$, for $w_{\phi0}=-0.7$, $-0.8$, and $-0.9$. The density parameter is fixed at $\Omega_{\phi0}=0.7$. The horizontal grey dashed line denotes $w=-1$.}
  \label{fig:w_z_wphi0}
\end{figure}

In this section, we derive the time evolution of $w_\phi$. In Ref.~\cite{Scherrer:2007pu}, the time dependence of $w_\phi$ was derived up to order $\lambda^2$ by evaluating $f_2(\Omega_\phi)$ using the $\Lambda\mathrm{CDM}$ expression for $\Omega_\phi$ (i.e., $\Omega_\phi$ is approximated by $\Omega_\Lambda$ in leading order). To consider higher order terms, we have to consider the correction in $\Omega_\phi$ itself. Therefore, in this section, we expand $\Omega_\phi$ itself in powers of $\lambda$:
\begin{align}
    \Omega_\phi&=\sum_{n=0}^{\infty}\Omega_\phi^{(n)}\lambda^n
    \label{expansion_of_Omega}
\end{align}
and derive $w_\phi$ as a function of redshift, up to order $\lambda^4$. Substituting Eq.~\eqref{expansion_of_Omega} into Eq.~\eqref{eq:domega_dlna} and equating the coefficients at each order in $\lambda$, we obtain
\begin{align}
    \lambda^0:\quad  \frac{d}{d\ln a}\Omega_\phi^{(0)}&=3\Omega_\phi^{(0)}(1-\Omega_\phi^{(0)})
    \label{eq:Omega_phi0}\\
    \lambda^1:\quad  \frac{d}{d\ln a}\Omega_\phi^{(1)}&=3\Omega_\phi^{(1)}(1-2\Omega_\phi^{(0)})\\
    \lambda^2:\quad  \frac{d}{d\ln a}\Omega_\phi^{(2)}&=3\Omega_\phi^{(2)}(1-2\Omega_\phi^{(0)})-3f_2(\Omega_\phi^{(0)})\Omega_\phi^{(0)}(1-\Omega_\phi^{(0)})-3\left\{ \Omega_\phi^{(1)} \right\}^2\\
    \lambda^3:\quad  \frac{d}{d\ln a}\Omega_\phi^{(3)}&=3\Omega_\phi^{(3)}(1-2\Omega_\phi^{(0)})\notag\\
    &\qquad-3\Omega_\phi^{(1)} \left\{ 2\Omega_\phi^{(2)}+f_2(\Omega_\phi^{(0)})\left( 1-2\Omega_\phi^{(0)} \right)\right.\notag\\
    &\qquad\qquad\qquad\qquad\qquad\qquad\left.+f_2'(\Omega_\phi^{(0)})\Omega_\phi^{(0)}\left( 1-\Omega_\phi^{(0)} \right)  \right\}
\end{align}
where a prime on $f_2$ denotes differentiation with respect to its argument. Solving the zeroth-order equation yields 
\begin{align}
    \frac{\Omega_\phi^{(0)}}{1-\Omega_\phi^{(0)}}=A_0a^3.
\end{align}
where $A_0$ is an integration constant. Its value depends on the choice of boundary conditions. We impose
\begin{align}
    \Omega_{\phi0}\equiv \Omega_\phi(a=1)=\Omega_\phi^{(0)}(a=1), \quad\Omega_\phi^{(i)}(a=1)=0\quad (i\geq1).
    \label{eq:boundary_condition}
\end{align}
This boundary condition yields $A_0=\Omega_{\phi0}/(1-\Omega_{\phi0})$ and hence
\begin{align}
    \Omega_\phi^{(0)}=\frac{1}{1+(\Omega_{\phi0}^{-1}-1)a^{-3}}.
\end{align}
This has the same form as the evolution of $\Omega_\Lambda$ in the $\Lambda\mathrm{CDM}$. Next, we consider the equations at first order in $\lambda$ and higher. First, for all orders equal to or greater than $1$, the homogeneous part of the equation takes the same form, namely, $\frac{d}{d\ln a}\Omega_\phi^{(n)}=3\Omega_\phi^{(n)}(1-2\Omega_\phi^{(0)})$. Solving the equation for $\Omega_\phi^{(1)}$ and integrating it gives 
\begin{align}
    \Omega_\phi^{(1)}&=A_1\frac{a^3}{\left( a^3+(\Omega_{\phi0}^{-1}-1) \right)^2},
\end{align}
where $A_1$ is an integration constant. From the boundary condition Eq.~\eqref{eq:boundary_condition}, it follows that
\begin{align}
    A_1=0.
\end{align}
Therefore, under the condition Eq.~\eqref{eq:boundary_condition}, we obtain
\begin{align}
    \Omega_{\phi}^{(1)}=0.
\end{align}
For odd-order equations of order $2n+1$, assuming that all odd-order terms up to order $2n-1$ vanish, 
the equation reduces to the homogeneous equation $\frac{d}{d\ln a}\Omega_\phi^{(2n+1)}=3\Omega_\phi^{(2n+1)}(1-2\Omega_\phi^{(0)})$. From the boundary condition $\Omega_\phi^{(2n+1)}(a=1)=0$, the solution for this homogeneous equation must be $0$. Therefore, by mathematical induction, we find
\begin{align}
    \Omega_{\phi}^{(2n+1)}=0 \quad (n=0,1,2,\cdots),
    \label{eq:odd_omega}
\end{align}
under the condition Eq.~\eqref{eq:boundary_condition}. Next, using Eq.~\eqref{eq:odd_omega}, the equation for $\lambda^2$ becomes
\begin{align}
    \frac{d}{d\ln a}\Omega_\phi^{(2)}=3\Omega_\phi^{(2)}(1-2\Omega_\phi^{(0)})-3f_2(\Omega_\phi^{(0)})\Omega_\phi^{(0)}(1-\Omega_\phi^{(0)}).
    \label{eq:Omega_phi2}
\end{align}
The solution for Eq.~\eqref{eq:Omega_phi2} can be written in the following integral form,
\begin{align}
    \Omega_\phi^{(2)}=-\Omega_\phi^{(0)}(1-\Omega_\phi^{(0)})\int\frac{f_2(\Omega_\phi^{(0)})}{\Omega_\phi^{(0)}(1-\Omega_\phi^{(0)})}d\Omega_\phi^{(0)},
\end{align}
where we used Eq.~\eqref{eq:Omega_phi0} to change the integration variable from $\ln a$ to $\Omega_\phi^{(0)}$. The integration can be performed analytically,
\begin{align}
    \int\frac{f_2(\Omega_\phi^{(0)})}{\Omega_\phi^{(0)}(1-\Omega_\phi^{(0)})}d\Omega_\phi^{(0)}
    &=-\frac{1}{2}f_2(\Omega_\phi^{(0)})+\frac{1}{3\sqrt{\Omega_\phi^{(0)}}}\ln\left( \frac{1+\sqrt{\Omega_\phi^{(0)}}}{1-\sqrt{\Omega_\phi^{(0)}}} \right)+C,
\end{align}
where $C$ is an integration constant. Therefore, under the boundary condition Eq.~\eqref{eq:boundary_condition}, Eq.~\eqref{eq:Omega_phi2} has the following solution:
\begin{align}
    \Omega_\phi^{(2)}=&-\Omega_\phi^{(0)}(1-\Omega_\phi^{(0)})\left\{ \frac{1}{2}f_2(\Omega_{\phi0})-\frac{1}{2}f_2(\Omega_\phi^{(0)})\right.\notag\\
    &\qquad\left.-\frac{1}{3\sqrt{\Omega_{\phi0}}}\ln\left( \frac{1+\sqrt{\Omega_{\phi0}}}{1-\sqrt{\Omega_{\phi0}}} \right)+\frac{1}{3\sqrt{\Omega_\phi^{(0)}}}\ln\left( \frac{1+\sqrt{\Omega_\phi^{(0)}}}{1-\sqrt{\Omega_\phi^{(0)}}} \right) \right\}.
\end{align}
From the above calculations, we can express $w_\phi$ as a function of redshift (scale factor) as follows:
\begin{align}
    w_\phi(z,\lambda,\Omega_{\phi0})&=-1+f_2(\Omega_\phi)\lambda^2+f_4(\Omega_\phi)\lambda^4+\mathcal{O}(\lambda^6),\notag\\
    &=-1+f_2(\Omega_\phi^{(0)}+\Omega_\phi^{(2)}\lambda^2+\cdots)\lambda^2+f_4(\Omega_\phi^{(0)}+\Omega_\phi^{(2)}\lambda^2+\cdots)\lambda^4+\mathcal{O}(\lambda^6),\notag\\
    &=-1+f_2(\Omega_\phi^{(0)})\lambda^2+\left\{ f_2'(\Omega_\phi^{(0)})\Omega_\phi^{(2)}+f_4(\Omega_\phi^{(0)}) \right\}\lambda^4+\mathcal{O}(\lambda^6).
    \label{eq:w_z_lambda_Omega_phi0}
\end{align}
For convenience, all the quantities appearing in the final expression are collected below.
\begin{align}
    \Omega_\phi^{(0)}&=\frac{1}{1+(\Omega_{\phi0}^{-1}-1)(1+z)^3}\\
    \Omega_\phi^{(2)}&=-\Omega_\phi^{(0)}(1-\Omega_\phi^{(0)})\left\{ \frac{1}{2}f_2(\Omega_{\phi0})-\frac{1}{2}f_2(\Omega_\phi^{(0)})\right.\notag\\
    &\qquad\left.-\frac{1}{3\sqrt{\Omega_{\phi0}}}\ln\left( \frac{1+\sqrt{\Omega_{\phi0}}}{1-\sqrt{\Omega_{\phi0}}} \right)+\frac{1}{3\sqrt{\Omega_\phi^{(0)}}}\ln\left( \frac{1+\sqrt{\Omega_\phi^{(0)}}}{1-\sqrt{\Omega_\phi^{(0)}}} \right) \right\}\\
    f_2(\Omega)&=\frac{1}{3}\left\{ \frac{1}{\sqrt{\Omega}}-\frac{1}{2}\left( \frac{1-\Omega}{\Omega} \right)\ln\left( \frac{1+\sqrt{\Omega}}{1-\sqrt{\Omega}} \right) \right\}^2\\
    f_4(\Omega)&=\frac{1-\Omega}{9\Omega}\left\{ \frac{1}{\sqrt{\Omega}}-\frac{1}{2}\left( \frac{1-\Omega}{\Omega} \right)\ln\left( \frac{1+\sqrt{\Omega}}{1-\sqrt{\Omega}} \right) \right\}\notag\\
    &\qquad\times\left\{ \frac{1}{2\sqrt{\Omega}}-\frac{3+\Omega}{4\Omega}\ln\left( \frac{1+\sqrt{\Omega}}{1-\sqrt{\Omega}} \right)\right.\notag\\
    &\qquad\qquad\left.+\frac{3-\Omega}{8\Omega^{3/2}}\ln^2\left( \frac{1+\sqrt{\Omega}}{1-\sqrt{\Omega}} \right)-\frac{(1-\Omega)^2}{16\Omega^2}\ln^3\left( \frac{1+\sqrt{\Omega}}{1-\sqrt{\Omega}} \right)\right\} .
\end{align}
Since Eq.~\eqref{eq:w_z_lambda_Omega_phi0} depends on $\lambda$, we consider replacing $\lambda$ with the current equation of state parameter of quintessence $w_{\phi0}$. In the expansion up to order $\lambda^4$, we have
\begin{align}
    1+w_{\phi0}\simeq f_2\left( \Omega_{\phi0} \right)\lambda^2+f_4\left( \Omega_{\phi0} \right)\lambda^4 .
\end{align}
Solving this for $\lambda^2$, we obtain
\begin{align}
    \lambda^2&=\frac{-f_2(\Omega_{\phi0})+\sqrt{f_2(\Omega_{\phi0})^2+4(1+w_{\phi0})f_4(\Omega_{\phi0})}}{2f_4(\Omega_{\phi0})},\notag\\
    &=\frac{1}{f_2(\Omega_{\phi0})}(1+w_{\phi0})-\frac{f_4(\Omega_{\phi0})}{f_2^3(\Omega_{\phi0})}(1+w_{\phi0})^2+\mathcal{O}\left( (1+w_{\phi0})^3 \right).
\end{align}
Here, we used $\lambda^2>0$ and the Maclaurin expansion of the square root with respect to $(1+w_{\phi0})$ up to the second order. Using this relation to eliminate $\lambda$ from Eq.~\eqref{eq:w_z_lambda_Omega_phi0}, we find $w_\phi$ as a function of $w_{\phi0}$ up to the second order in $1+w_{\phi0}$:
\begin{align}
    1+w_\phi=&\frac{f_2(\Omega_{\phi}^{(0)})}{f_2(\Omega_{\phi0})}(1+w_{\phi0})\notag\\
    &\quad+\frac{1}{f_2(\Omega_{\phi0})^2}\left\{ -f_2(\Omega_\phi^{(0)})\frac{f_4(\Omega_{\phi0})}{f_2(\Omega_{\phi0})}+f_2'(\Omega_{\phi}^{(0)})\Omega_{\phi}^{(2)}\right.\notag\\
    &\left.\qquad\qquad\qquad\qquad+f_4(\Omega_\phi^{(0)}) \right\}(1+w_{\phi0})^2+\mathcal{O}\left( (1+w_{\phi0})^3 \right).
    \label{eq:w_z_w_phi0}
\end{align}
\footnote{
    Here, we clarify the difference between our result and previous works. In Ref.~\cite{Chiba:2009sj}, by adopting the approximation that the energy scale of dark energy is given by the potential energy at the initial value of the scalar field ($\rho_{\phi}\simeq \rho_{\phi0}\simeq V(\phi_i)$) and using the expression for the scale factor of the $\Lambda\mathrm{CDM}$ model, the following formula for $w$ was obtained:
    \begin{align}
        1+w(a)\simeq(1+w_0)a^{3(K-1)}\left(\frac{(K-F(a))(F(a)+1)^K+(K+F(a))(F(a)-1)^K}{(K-\Omega_{\phi 0}^{-1/2})(\Omega_{\phi 0}^{-1/2}+1)^K+(K+\Omega_{\phi 0}^{-1/2})(\Omega_{\phi 0}^{-1/2}-1)^K}\right)^2,
        \label{eq:chiba_2009}
    \end{align}
    where $K\equiv\sqrt{1-\frac{4}{3}\frac{V_{,\phi\phi}(\phi_i)}{V(\phi_i)}m_{\mathrm{pl}}^2}$ and $F(a)\equiv\sqrt{1+(\Omega_{\phi 0}^{-1}-1)a^{-3}}$. This expression was initially derived for hilltop quintessence in \cite{Dutta:2008qn}. For an exponential potential, $K=\sqrt{1-\frac{4}{3}\lambda^2}$. Expanding this expression in powers of $(1+w_{\phi0})$ successfully reproduces the linear coefficient of $(1+w_{\phi0})$ in Eq.~\eqref{eq:w_z_w_phi0}, but fails to match the second-order coefficient. This discrepancy arises from the assumptions made in Ref.~\cite{Chiba:2009sj}, namely the use of the $\Lambda\mathrm{CDM}$ scale factor and the approximation $\rho_{\phi}\simeq \rho_{\phi0}\simeq V(\phi_i)$.
}
In Fig.~\ref{fig:w_comparison}, we compare the numerical solution with the expansion up to the first and second orders in $(1+w_{\phi0})$. It can be seen that the expansion up to the second order in $(1+w_{\phi0})$ improves the accuracy of the approximation relative to the first-order result. For the case with $\lambda=1.0$ and $\Omega_{\phi0}=0.70$, the relative error in $1+w_\phi$ at $z=1$ decreases from $13.9\%$ for the leading-order approximation to $1.6\%$ when the next-order correction is included.
Furthermore, Figs.~\ref{fig:w_z_omega} and \ref{fig:w_z_wphi0} plot the approximate expression for $w_\phi(z)$ up to the second order in $(1+w_{\phi0})$ for different values of $\Omega_{\phi0}$ with a fixed $w_{\phi0}$, and for different values of $w_{\phi0}$ with a fixed $\Omega_{\phi0}$, respectively.

\section{Conclusion}\label{sec:conclusion}
In this study, we analytically expanded the thawing behavior of quintessence with an exponential potential. By expanding $w_\phi$ around $-1$ in terms of $\lambda$ as $w_\phi=-1+f_2(\Omega_\phi)\lambda^2+f_4(\Omega_\phi)\lambda^4$, we analytically derived the $\lambda^4$ coefficient $f_4(\Omega_\phi)$, in addition to the leading-order $f_2(\Omega_\phi)$ previously obtained in \cite{Scherrer:2007pu}. We also showed that by expanding these coefficients in terms of $\Omega_\phi$,  they agree with the expansion coefficients obtained when $w_\phi$ is expanded directly in $\Omega_\phi$. 

Furthermore, to determine the redshift dependence of $w_\phi$ up to $\mathcal{O}(\lambda^4)$ accurately, we expanded $\Omega_\phi$ itself in terms of $\lambda$. Under the boundary condition that the higher-order terms of $\Omega_\phi$ vanish at present and the lowest-order term coincides with $\Omega_{\phi0}$, we analytically obtained the first non-trivial correction, $\Omega_\phi^{(2)}$. Consequently, we found terms originating from the correction, such as $f_2'(\Omega_\phi)\Omega_{\phi}^{(2)}$ and derived the redshift dependence of $w_\phi$ up to $\mathcal{O}(\lambda^4)$ consistently.

Moreover, by comparing with numerical calculations, we confirmed that the approximate formula for the redshift dependence of $w_\phi$ up to $\lambda^4$ improves the accuracy compared to the expansion up to $\lambda^2$. For the case with $\lambda=1.0$ and $\Omega_{\phi0}=0.70$, including the next-order correction reduces the relative error in $1+w_\phi$ at $z=1$ from $13.9\%$ to $1.6\%$. Finally, we eliminated $\lambda$ using the present equation of state parameter $w_{\phi0}$ and derived an analytical expression representing $w_\phi(z)$ in terms of $\Omega_{\phi0}$ and $w_{\phi0}$. This expression could serve as a potentially useful tool when comparing with observational data.

\begin{acknowledgments}
This work was in part supported by JSPS KAKENHI Grant
No. JP24K07027 (KK).
\end{acknowledgments}

\appendix
\section{derivation of \texorpdfstring{$f_4$}{f4}}\label{sec:f_4_derivation}
Here, we derive the analytical solution of $f_4(\Omega_\phi)$ from Eq.~\eqref{eq:f_4_equation}. Furthermore, we obtain the series expansion representation Eq.~\eqref{eq:f_4_series_expansion} of $f_4(\Omega_\phi)$ with respect to $\Omega_\phi$. First, the differential equation Eq.~\eqref{eq:f_4_equation} for $f_4(\Omega_\phi)$ can be rearranged as follows:
\begin{align}
    \frac{df_4}{d\Omega_\phi}&=\frac{1}{3\Omega_\phi(1-\Omega_\phi)}\left( \sqrt{\frac{3\Omega_\phi}{f_2}}-6 \right)f_4+\frac{f_2^2}{\Omega_\phi(1-\Omega_\phi)}-\frac{f_2\sqrt{3\Omega_\phi f_2}}{3\Omega_\phi(1-\Omega_\phi)}+f_2\frac{df_2}{d\Omega_\phi}\notag\\
    &=\left\{ \frac{1}{\sqrt{3\Omega_\phi}(1-\Omega_\phi)}\frac{1}{\Theta}-\frac{2}{\Omega_\phi(1-\Omega_\phi)} \right\}f_4+\frac{1}{2}f_2\frac{df_2}{d\Omega_\phi}\notag\\
    &=\left\{ \frac{\Theta'}{\Theta}-\frac{1}{\Omega_\phi(1-\Omega_\phi)} \right\}f_4+\Theta^3\Theta'
\end{align}
This yields an integral representation of $f_4$. 
\begin{align}
    f_4(\Omega_\phi)=\frac{(1-\Omega_\phi)\sqrt{f_2}}{2\Omega_\phi}\int d\Omega_\phi\frac{\Omega_\phi}{(1-\Omega_\phi)}f_2'\sqrt{f_2}
\end{align}
As we will see later this form is particularly advantageous for deriving the power series expansion, but the integral itself can also be performed using only elementary functions. To perform integration, we first rewrite the integral as follows:
\begin{align}
   &\frac{1}{2}\int d\Omega_\phi\frac{\Omega_\phi}{1-\Omega_\phi}f_2'\sqrt{f_2}\notag\\
    &=\frac{1}{3}\frac{\Omega_\phi}{1-\Omega_\phi}f_2^{3/2}-\frac{1}{3}\int d\Omega_\phi\left( \frac{\Omega_\phi}{1-\Omega_\phi} \right)'f_2^{3/2},\notag\\
    &=\frac{1}{3}\frac{\Omega_\phi}{1-\Omega_\phi}f_2^{3/2}-\frac{1}{9\sqrt{3}}\int d\Omega_\phi\frac{1}{(1-\Omega_\phi)^2}\left\{ \frac{1}{\sqrt{\Omega_\phi}}-\frac{1}{2}\left( \frac{1}{\Omega_\phi}-1 \right)\ln\left( \frac{1+\sqrt{\Omega_\phi}}{1-\sqrt{\Omega_\phi}} \right) \right\}^3.
\end{align}
Here, defining $X=\sqrt{\Omega_\phi}$ and 
\begin{align}
    L=\ln\left( \frac{1+X}{1-X} \right),
\end{align}
the integral in the second term becomes
\begin{align}
    &\frac{1}{9\sqrt{3}}\int d\Omega_\phi\frac{1}{(1-\Omega_\phi)^2}\left\{ \frac{1}{\sqrt{\Omega_\phi}}-\frac{1}{2}\left( \frac{1}{\Omega_\phi}-1 \right)\ln\left( \frac{1+\sqrt{\Omega_\phi}}{1-\sqrt{\Omega_\phi}} \right) \right\}^3\notag\\
    &=\frac{2}{9\sqrt{3}}\int dX\left\{  -\frac{1-X^2}{8X^5}L^3+\frac{3}{4X^4}L^2-\frac{3}{2X^3(1-X^2)}L+\frac{1}{(1-X^2)^2X^2} \right\}.
\end{align}
Next, we separate the integral into four parts and define $I_1, I_2, I_3,$ and $I_4$ as follows:
\begin{align}
    I_1&\equiv\int\left( -\frac{1-X^2}{8X^5} \right)L^3dX,\\
    I_2&\equiv\int\frac{3}{4X^4}L^2dX,\\
    I_3&\equiv\int \left( -\frac{3}{2X^3(1-X^2)} \right)LdX,\\
    I_4&\equiv\int \frac{1}{X^2(1-X^2)^2}dX.
\end{align}
$I_1$ can be rearranged as 
\begin{align}
    I_1&=\frac{1-2X^2}{32X^4}L^3-\int\frac{3(1-2X^2)}{16X^4(1-X^2)}L^2dX
\end{align}
where we have used the relation $\frac{dL}{dX}=2/(1-X^2)$.
Adding the remaining integral of $I_1$ to $I_2$, we obtain
\begin{align}
    &I_2-\int\frac{3(1-2X^2)}{16X^4(1-X^2)}L^2dX\notag\\
    &=\frac{3}{16}\int \left\{\frac{1}{1-X^2} +\frac{3}{X^4}+\frac{1}{X^2} \right\}L^2dX,\notag\\
    &=\frac{3}{32}\int \frac{dL}{dX}L^2dX +\frac{3}{16}\int\left( \frac{3}{X^4}+\frac{1}{X^2} \right)L^2dX,\notag\\
    &=\frac{1}{32}L^3-\frac{3}{16}\frac{1+X^2}{X^3}L^2+\frac{3}{4}\int\frac{1+X^2}{X^3(1-X^2)}LdX.
\end{align}
By adding the remaining integral term to $I_3$, we find
\begin{align}
    &I_3+\frac{3}{4}\int\frac{1+X^2}{X^3(1-X^2)}LdX\notag\\
    &=\int \left( -\frac{3}{4X^3} \right)LdX,\notag\\
    &=\frac{3}{8}\left( \frac{1-X^2}{X^2} \right)L+\frac{3}{4X}.
\end{align}
For $I_4$, we can integrate it as
\begin{align}
    I_4&=-\frac{1}{4}\frac{1}{1+X}+\frac{1}{4}\frac{1}{1-X}+\frac{3}{4}L-\frac{1}{X}.
\end{align}
For simplicity, we omit the integration constant, which is
fixed to zero by the thawing boundary condition. Therefore, summing these four integrals gives
\begin{align}
    I_1+I_2+I_3+I_4&=\frac{(1-X^2)^2}{32X^4}L^3-\frac{3}{16}\frac{1+X^2}{X^3}L^2+\frac{3}{8}\frac{1+X^2}{X^2}L+\frac{3X^2-1}{4X(1-X^2)}.
\end{align}
Substituting this result back, we finally obtain
\begin{align}
    f_4&(\Omega_\phi)=\frac{(1-\Omega_\phi)\sqrt{f_2}}{2\Omega_\phi}\int d\Omega_\phi\frac{\Omega_\phi}{1-\Omega_\phi}\sqrt{f_2}f_2'\notag\\
    &=\frac{1-\Omega_\phi}{9\Omega_\phi}\left\{ \frac{1}{\sqrt{\Omega_\phi}}-\frac{1}{2}\left( \frac{1-\Omega_\phi}{\Omega_\phi} \right)\ln\left( \frac{1+\sqrt{\Omega_\phi}}{1-\sqrt{\Omega_\phi}} \right) \right\}\notag\\
    &\qquad\times\left\{ \frac{1}{2\sqrt{\Omega_\phi}}-\frac{3+\Omega_\phi}{4\Omega_\phi}\ln\left( \frac{1+\sqrt{\Omega_\phi}}{1-\sqrt{\Omega_\phi}} \right)\right.\notag\\
    &\qquad\qquad\left.+\frac{3-\Omega_\phi}{8\Omega_\phi^{3/2}}\ln^2\left( \frac{1+\sqrt{\Omega_\phi}}{1-\sqrt{\Omega_\phi}} \right)-\frac{(1-\Omega_\phi)^2}{16\Omega_\phi^2}\ln^3\left( \frac{1+\sqrt{\Omega_\phi}}{1-\sqrt{\Omega_\phi}} \right)\right\} .
\end{align}

\bibliography{reference}
\end{document}